\documentclass[a4paper,12pt]{article}
 \usepackage{graphicx}
 \usepackage{epsfig}
\usepackage{amsmath}
\usepackage{amssymb}
\newcommand{\sech}{\mbox{\rm sech}}
\newcommand{\cosech}{\mbox{\rm cosech}}
\usepackage[mathscr]{eucal}

\newcommand{\dd}[2][]{\frac{d#1}{d#2}}
\newcommand{\DD}[2][]{\frac{d^2 #1}{d#2^2}}
\newcommand{\Hsc}{\mathscr{H}}
\title
{\bf Shape Invariant Potentials in Higher Dimensions}

\author{
 R. Sandhya$^{a}$, S. Sree Ranjani$^{b,c}$ and A. K. Kapoor$^{d,e}$\\
$^a$Department of Physics, Osmania University, Hyderabad, India. \\[3mm]
$^b$School of Physics, University of Hyderabad,\\ Hyderabad, India, 500
046.\\[3mm]
$^{c}$Faculty of Science of Technology,\\ ICFAI foundation for Higher
Education,\\
(Declared as deemed-to-be University u/s 3 of the UGC Act 1956),\\
Dontanapally, Hyderabad, India, 501203 \footnote{ Present
Address}$^*$.\\[3mm]
$^d$Department of Physics, Shiv Nadar University,\\
Village Chithera, Tehsil Dadri, Gautam Budh Nagar,\\
UP, India, 203207.\\[3mm]
$^{e}$ Department of Physics, BITS-Pilani Hyderabad Campus\\
Jawahar Nagar, Shameerpet Mandal,\\
Hyderabad 500078, India.$^*$
}

\begin{document}

\maketitle

\begin{abstract}
In this paper we investigate the shape
invariance property of a potential in one dimension. We show that a simple
ansatz allows us to reconstruct all the known shape invariant potentials in one 
dimension. This ansatz can be easily extended to arrive at a large class of new
shape invariant potentials in arbitrary dimensions.
A reformulation of the shape invariance property and possible generalizations
are proposed. These may lead to an important extension of the shape invariance
property to Hamiltonians that are related to standard potential problems via space
time transformations, which are found useful in path integral formulation of quantum mechanics. 
\end{abstract}

\section{Introduction}
   Supersymmetric quantum mechanics (SUSYQM) \cite{witten} played an important
role in understanding the problem of solvable potentials. The shape invariance
(SI) condition introduced by Gendenshtein \cite{gad} gives a sufficient
condition for the solvability of a given one dimensional potential. This helped
in obtaining a whole class of shape invariant potentials (SIP) which are
analytically exactly solvable (ES). It also gave useful insights into the
factorization method \cite{infeld} and has played an important role in the past
three decades in constructing new ES potentials in one
dimension \cite{khare_book},
\cite{phys_rep}.  More recently supersymmetry (SUSY) has provided another
route to the construction of  rational potentials \cite{quesne},
\cite{ces}, which have the new exceptional orthogonal polynomials (EOPs) as
a part of their solutions \cite{kam}, \cite{kam_nick}.
	
In higher dimensions the number of known solvable potential models,
other than separable ones is severely restricted \cite{Ioffe} - \cite{Cannata}.
The power of SUSY and the SI condition give a hope that one may
construct a new class of ES models in higher dimensions. At present, work on
SI in higher dimensional models is very limited. Use of the SI
requirement has been shown to lead to known  ES  potential models in one
dimension \cite{asim}.

In this paper, we present, an alternate route  to
analyse and obtain the solutions to SI requirement.
This is achieved by  the use of a simple ansatz which leads to the solution
for the superpotential in terms of the free particle  Schr\"odinger
equation solutions.  This approach and its extension reproduces all the known SI
potentials in one dimension, including those recently discovered potentials related to EOPs
\cite{quesne} -\cite{kam_nick}. In addition,  this  route to the solutions of the SI
requirement trivially generalizes to higher dimensions and leads to, in
the first instance, a large class of SI potentials which get related to the
solutions of free particle Schr\"odinger equation in higher dimension.

Even in one dimension, the SI property alone is not sufficient for obtaining
exact solutions for the bound state energy eigenvalues and eigenfunctions. In
addition one needs to use the intertwining property of partner Hamiltonians.
For separable potential models, the intertwining property  works
exactly in the same way as in one dimension, as the solution of the problem
reduces to solution of several one dimensional problems. The known examples of
intertwining property in two dimensions indicate that, in practice, this 
property may have a limited role to play in higher dimensional
models and one needs a fresh approach.

A possible route may be the use of space time transformations found useful
for obtaining exact solutions of quantum mechanical problems in the path
integral approach. In 1984, Duru and Kleinert used space time transformations to
provide an exact solution of hydrogen atom problem in three dimensions within
the path integral formalism \cite{Duru}, \cite{Duru1}.  In the next ten years it
was shown to be useful for obtaining exact path integral solutions of many other
problems\cite{Ho} -\cite{HKlein}. Further discussion of space time
transformations is given in \cite{ST1}-\cite{ST4}. It is important to mention
that the use of space time transformations is not limited to the path integral
formalism alone and have been found to be useful beyond one dimensional
potential problems. Inspired by the success of the space time transformations,
a generalization of shape invariance requirements to Hamiltonians of
a more general form is presented. As an example, it is applied to the radial
equation for free particle equation in three dimensions to show that this leads to known
recurrence relations between spherical Bessel functions.

The paper is organized as follows. In the next section we briefly
describe SUSYQM and we give an alternate definition of the SI
condition and follow a new approach to obtain solutions of SI requirement. It
is shown that all the known ES solvable potentials in one dimension are
obtained by making use of few simple ansatz. In section 3, we
show that, in arbitrary dimensions, this route to the analysis of SI condition does
not require anything new and easily leads to SIPs. In section 4, we summarize our results and 
conclude by giving routes to further generalizations of the SI property.

\noindent
\section{Supersymmetric quantum mechanics }
       In SUSYQM \cite{witten}, \cite{phys_rep}, we have a pair of partner
potentials $V^{\pm}(x)$ defined in terms of the suprepotential $W(x)$ as
\begin{equation}
V^{(\pm)}(x)= W^2(x) \pm W^{\prime}(x),  \label{partners}
\end{equation}					
where the prime denotes differentiation with respect to $x$ and  $W(x)$ is
defined as
\begin{equation}
W(x)= - \frac{d}{dx}\log \psi_0^{(-)}(x). \label{suppot}
\end{equation}
Here $\psi_0^{(-)}(x)$ is the ground state wave function of $V^-(x)$ and
$E_0^-=0$.
The wave functions $\psi^{(\pm)}_n(x)$ of the partners are related by
\begin{equation}
\psi_{n+1}^{(-)}(x)=A^{\dag}\psi_n^{(+)}(x)  \,\,\,;\,\,\, \psi_n^{(+)}(x)=A
\psi_{n+1}^{(-)}(x),  \label{wfs}
\end{equation}
 where $A$ and $A^{\dag}$ are the intertwining operators
\begin{equation}
A=\frac{d}{dx}+W(x)\,\,\,;\,\,\,A^{\dag}=-\frac{d}{dx}+W(x). \label{intertwine}
\end{equation}
SUSY is known to be unbroken (exact) between the partners $V^{(\pm)}(x)$,  if
$E_0^-=0$, $\psi_0^{(-)}$ is  normalizable and $A\psi^{(-)}_0=0$. In this case,
$\psi_0^{(+)}(x)$ is non-normalisable and the partners are isospectral except
for the ground states. SUSY is said to be broken, when the ground states of both
the
partners are non-normalisable and the partners are isospectral including the
ground states. For more details we refer the reader to \cite{phys_rep},
\cite{khare_book}.\\

It should be noted that for a given potential $V^{(-)}(x)$, superpotential
$W(x)$ is not uniquely determined. {\it Moreover, each $W(x)$
associated with $V^{(-)}(x)$, in  general, gives
a different partner associated with it. Therefore, it would appear that
the shape invariance property depends on the choice of the superpotential; this
statement is not correct.} In the next section a sufficient condition for shape
invariance is formulated in terms of the solutions of quantum Hamilton-Jacobi (QHJ) equation \cite{lea1}, \cite{sree}. This  condition makes no
reference to any particular choice of the superpotential.

\section*{Shape invariance}
The SUSY partners are said to be shape invariant if, for some $f(\lambda)$, one
has
\begin{equation}
V^{(+)}(x,\lambda)= V^{(-)}(x, f(\lambda)) + R(\lambda),\label{SI}
\end{equation}
where $\lambda$ is the potential parameter and $R(\lambda)$ is a function of
$\lambda$. It is clear that given a potential $V(x, \lambda)$, depending on a set
of parameters $\lambda$, one introduces a superpotential $W(x, \lambda)$ such
that
\begin{equation}
 V(x) = W^2(x,\lambda) - W^{\prime}(x,\lambda) + E_0,  \label{EQ012}
\end{equation}
which has the form of Riccati equation and is also known as the
QHJ equation . Substituting \eqref{suppot} in the above equation
gives the Schr\"odinger equation. It may be remarked that several solutions of
the above equation for $W(x)$
exist. For a mapping of parameters $ \rho : \lambda \to
\rho(\lambda)$, we introduce functions $w_1, w_2$ through equations
\begin{eqnarray}
w_1(x, \lambda) = -W(x,\lambda), \qquad \text{and}\quad
w_2(x,\lambda) = W(x, \tau(\lambda)),
\end{eqnarray}
and define two potentials $V_1,V_2$ by
\begin{eqnarray}
&& V_k = w_k^2(x,\lambda) - w^{\prime}_k(x, \lambda)), \quad k=1, 2.
\end{eqnarray}
A potential $V(x)$ is called shape invariant, if one can find a superpotential
$W(x)$ and a mapping $\lambda \to \tau(\lambda) $ such that
$V_1(x,\lambda)$ and $V_2(x,\lambda)$ differ by a constant. Though the use of
partner potential has been bypassed here, it is apparent that this definition
of shape invariance is equivalent to that used in literature.

It is to be noted that an obvious sufficient condition for shape invariance of a
potential
$V(x,\lambda)$ is
the existence of superpotential $W(x)$ and a map $\tau$ such that
\begin{equation}
  W(x, \tau(\lambda))= -W(x,\lambda).  \label{EQ10}
\end{equation}
The above sufficient condition for shape invariance can be restated as follows.
A potential $V(x,\lambda)$ is shape invariant if there exists a superpotential
$W(x,\lambda)$ satisfying the QHJ equation
\begin{equation}
 W^2(\lambda,x) -W^\prime(\lambda,x)= V(x,\lambda) -E_1
\end{equation}
and a map $\tau$ such that the function $\underline{W}$ defined by
\begin{equation}
\underline{W}(\lambda, x)\equiv -W(\tau(\lambda),x),
\end{equation}
satisfies the QHJ equation
\begin{equation}
  \underline{W}^2(x,\lambda) - \underline{W}^\prime(x,\lambda)= V(x) -E_2,
\end{equation}
for some energy $E_2$.

In the next section we re-derive the one dimensional SIPs listed in
\cite{khare_book}, in a very simple and straightforward fashion, by making
a simple ansatz for the superpotential. Our 
approach has  the advantage of  being very simple and also leads one to a large 
class of shape invariant potentials in higher dimensions.\\

\section*{A simple derivation of SI potentials in one dimension}
In the table below, we list the known SIPs in one dimension along with the corresponding  superpotentials \cite{khare_book}, \cite{phys_rep}. 

\vspace{5mm}
\begin{tabular}{|l|l|l|l|}
\hline&&&\\
Name of Potential & Superpotential & Name of Potential & Superpotential \\[2mm]
\hline &&&\\
 Shifted Oscillator & $\dfrac{1}{2}{}\omega x -b$ &
 Radial Oscillator & $\dfrac{1}{2}{}\omega r - \dfrac{\ell+1}{r}$\\[3mm]
 Coulomb & $\dfrac{e^2}{2(\ell+1)} - \dfrac{\ell +1 }{r}$ &
 Morse   & $A - B \exp(-\,a{}x)$ \\[3mm]
 Scarf II & $A\tanh \,a{}x + B \sech \,a{}x$ &
 Rosen-Morse II & $A\tanh \,a{}x + B/A$ \\
  ~~~~(hyperbolic)&& ~~~~(hyperbolic) & \\[3mm]
 Eckart & $-A\coth \,a{}r + B/A$ &
 Scarf I & -$A \tan a x + B \sec ax$ \\
 && ~~~(trignometric) & \\[3mm]
 Gen. P\"{o}schl-Teller & $A \coth\, ar -B \cosech\, ar $ &
 Rosen-Morse & $-A \cot\, ax - B/A$ \\ &&~~~~trignometric& \\
\hline
\end{tabular}
\vspace{5mm}

We will now present a simple derivation of all these SIPs. We will show in the next section how
the simplicity of this derivation makes it possible to generalize the results to higher dimensions.\\[3mm]
Let the SUSY partner potentials, $V^{(\pm)}(x,\lambda)$, be specified by a superpotential $W(x)$:
\begin{equation}
    V^{(\pm)}(x,\lambda) = W^2(x,\lambda) \pm W^\prime(x,\lambda) \label{eq31},
\end{equation}
 where $\lambda$ denotes a parameter appearing in the potential.

On  using a simple ansatz
\begin{equation}
  W(x,\lambda)=\lambda F(x), \label{WequaltoLF}
\end{equation}
the shape invariance requirement \eqref{eq31} becomes
\begin{equation}
   (\lambda^2 - \mu^2) F(x)^2 + (\lambda+\mu) F^\prime(x) + R =0 \label{EQ031}.
\end{equation}
Defining  a scaled independent variable $(\lambda-\mu)x\equiv \xi$, we get
\begin{equation}
     \tilde{F}^2(\xi) + \tilde{F}^\prime(\xi) + K =0,  \label{1EQ04}
\end{equation}
where $K=R/(\mu+\lambda)$ is a constant and $\tilde{F}(\xi)= F(x(\xi))$.
This equation has the form of the Riccati equation and can be linearised by
introducing a function, $u(\xi)$, by $\tilde{F}(\xi) \equiv
u^\prime(\xi)/u(\xi)$. The equation for $u(\xi)$ turns out to be the familiar free particle Schr\"{o}dinger
equation:
\begin{equation}
     u^{\prime\prime}(\xi) + K u(\xi) =0, \label{1EQ05}
\end{equation}
whose solutions can be written down immediately for different cases  $K>0,
K=0$ and $K<0$. Thus for
each case of $K$, we use the corresponding solution of \eqref{1EQ05} and
calculate $W(x)$ as follows.

\paragraph*{Case I $K=0$ :}
\begin{eqnarray}
   u(\xi) &=&A\xi + B, \qquad W(x) =\lambda \frac{A}{A(\alpha x) + B } \label{new}\\
  \text{where}\ \ \  \alpha &=& (\lambda-\mu) \label{1EQ06}
\end{eqnarray}
and $A,B$ are constants.

\paragraph*{Case II $K >0$:}
Using the notation $k=\sqrt{K}$ the two independent solutions are
\begin{eqnarray}
     u(\xi) = \sin \xi \,\,,\,\, \cos\xi \label{EQ07}  \\  
     W(x) =\lambda k \tan (k \alpha x), - \lambda k \cot (k \alpha x)
\label{1EQ08}.
\end{eqnarray}

\paragraph*{Case III $K <0$:}
In this case, the two independent solutions are given in terms of the hyperbolic 
functions and
\begin{eqnarray}
     u(\xi) = \sinh c \xi \,\,,\,\, \cosh c\xi \label{EQ09}  \\  
     W(x) =\lambda c \tan( c \alpha x )\,,\, \lambda c \cot( c \alpha x)
\label{1EQ10}.
\end{eqnarray}
where $c^2=-K$.

The superpotentials constructed above correspond to the one parameter potentials
like the harmonic oscillator, trigonometric Scarf, Rosen-Morse, hyperbolic
P\"osch-Teller etc., with suitable redefinition of parameters, listed in the table.

\subsubsection*{Extensions}
As a next step we wish to extend the above results to cover the other cases
listed in the table.  We will construct superpotentials corresponding to the two parameter families of potentials. We will
use the fact that a solution of Riccati equation can be used to 
construct a second solution \cite{piaggio}. Denoting a solution of \eqref{1EQ04}
by $f(\xi)$
and writing  $ W(\xi) = \lambda f(\xi) +\phi(\xi)$,  assuming $\phi(\xi)$ to be 
independent of $\lambda$, an analysis of the shape 
invariance requirement leads to 
\begin{equation}
 f (\xi) \phi(\xi)  + \phi^\prime(\xi) =C,
\end{equation}
where $C$ is a constant. The solution for $\phi(\xi)$ in terms of $f(\xi)$ is given by
\begin{equation}
   \phi(\xi) = e^{-  \int f(\xi)\,d\xi}\left( C e^{\int f(\xi)\,d\xi} +D 
\right),  
\end{equation}
where $D$ is the constant of integration. Thus each of the 
above solutions, given in \eqref{new}, \eqref{1EQ08} and \eqref{1EQ10}, can be
used to construct  a new solution of \eqref{1EQ04}. At this stage, we get the radial oscillator, Scarf I and II, generalized
P\"{o}schl-Teller and $B=0$ cases of the Rosen Morse-I, Rosen-Morse-II and the Eckart
potentials.

For the remaining potentials, {\it i.e.}, the shifted oscillator, Coulomb, Morse, Eckart
Rosen-Morse-I and II  potentials with $B \ne 0$, we proceed as
follows. Let $F(x)$ be one of the solutions as obtained above. We now look for
solutions of the shape invariance condition of the form
\begin{equation}
      W(x) =\lambda F(x) + g(\lambda,x). \label{EQ11} 
\end{equation}
The general case, when $g$ depends on both $x$ and $\lambda$ is complicated to 
analyse. For the present purpose, it turns out to be sufficient to assume that
$g$ is independent of $x$.  In this case
we get
\begin{eqnarray}
 \Big(\lambda F(x) + g(\lambda) \Big)^2 + \lambda F^\prime(x)  
 -\Big(\mu F(x) + g(\mu)\Big)^2 + \mu F^\prime (x) + R =0.  \label{EQ111}
\end{eqnarray}
Use of \eqref{EQ031} and the fact that $F(x)$  is one of the solutions already
giving a shape invariant potential, simplifies the above equation to give
\begin{equation}
    g(\lambda) =\text{const}/\lambda.
\end{equation}
This completes the construction of all the superpotentials listed in the table 
and we can reproduce all the known SIPs in one dimension with suitable choice 
of the constants.

Let us assume that we have found a set of solutions of the SI requirement
\eqref{EQ111}. We now ask if we can extend these solutions to further generate a
new set of SIPs? Let $W(x,\lambda)$ be a superpotential corresponding to a shape invariant potential. 
Let us introduce $\tilde{W}(x,\lambda)= W(x,\lambda) + \chi(x,\lambda)$  and the requirement that the
corresponding potential $\tilde{V}(x,\lambda)$ being shape invariant gives
\begin{equation}
 \tilde{W}(x,\lambda)^2 - \tilde{W}^\prime(x,\lambda) = \tilde{W}(x,\mu)^2 +
           \tilde{W}^\prime(x,\mu) + R(\mu,\lambda),
\end{equation}
where $R(\mu,\lambda)$ is a constant. Substituting for $\tilde{W}(x,\lambda)$ and making 
use of the shape invariance property of $W(x,\lambda)$ and simplifying we get
\begin{eqnarray}
  \lefteqn{\chi(x,\lambda)^2 + 2 W(x,\lambda) \chi(x,\lambda) +
\chi^\prime(x,\lambda)}\\
   &&=\chi(x,\mu)^2 + 2 W(x,\mu) \chi(x,\mu) - \chi^\prime(x,\mu)
     +  R(\mu,\lambda). \label{EQ29A}
\end{eqnarray}
Solutions of the above equation, for the case when $\chi(x,\lambda)$ is a
constant have already been found.
The general case where $\chi(x,\lambda)$ is a function of $x$, does not appear
to be easy to solve. 
%
Recall for shape invariance to be satisfied, there exists a mapping $\tau$ such
that $W(x,\tau(\lambda)) = - W(x,\lambda)$. The above equation can then be cast
 in the form
 \begin{eqnarray}
  \lefteqn{\chi(x,\lambda)^2 + 2 W(x,\lambda) \chi(x,\lambda) +
\chi^\prime(x,\lambda)}\\
   &&=\chi(x,\tau(\mu))^2 + 2 W(x,\tau(\mu)) \chi(x,\tau(\mu)) +
\chi^\prime(x,\tau(\mu))
     +  R(\mu,\lambda). \label{EQ29e}
\end{eqnarray}
One can make further progress by looking for solutions that satisfy the ansatz
 \begin{equation}
 \chi(x,\lambda)^2 + 2 W(x,\lambda) \chi(x,\lambda) +\chi^\prime(x,\lambda)
 = K(\lambda),
\end{equation}
as suggested by the above equation \eqref{EQ29e}. Here $K(\lambda)$ is a constant.  Among other results, we are
then led to the recently discovered SIPs with solutions related to EOPs \cite{quesne}, \cite{sasaki}, \cite{kam},
\cite{kam_nick}.

This ansatz turns out to be equivalent to an isospectral shift deformation of
the original potential, first  used in \cite{ces} for the construction of new rational potentials with EOPs as solutions. 
In this work it was not clear if the extended potentials were shape invariant, in fact it was erroneously concluded that the
deformed potentials are not shape invariant. Our approach with SI as guiding principle guarantees that the deformed potentials indeed
turn out to be shape invariant. The simple process outlined in this section allows us to construct these new potentials from the
requirements of shape invariance. The details of this study are reported elsewhere \cite{isoshift}.

%

In the next section, we show how our approach to SI not only
allows us to extend the concept of SI to higher dimensions but
also provide us with a method to construct SI potentials in higher dimensions.

\noindent
\section{\bf SI in higher dimensions}

In this section, we reformulate the shape invariance property in one
dimension in a manner which can generalized to higher dimensions. To construct
SIPs in higher dimensions is straight forward. For this it turns out to be useful and simpler to discard the 
superpotential and  to write the Hamitonian in a different form.  To elucidate,
we consider the one
dimensional Hamiltonian
\begin{equation}
     H = p^2 + W^2(x,\lambda) - W^\prime(x,\lambda),
\end{equation}
which can be rewritten as
\begin{equation}
     H = - e^{\Omega(x)}\dd{x} e^{-2\Omega(x)} \dd{x} e^{\Omega(x)},\label{EQ01}
\end{equation}
where $\Omega(x,\lambda)$, called prepotential, is related to the
superpotential by
\begin{equation}
    \Omega(x,\lambda) = \int W(x,\lambda) \, dx .\label{EQ021}
\end{equation}
It is easy to see that 
\begin{equation}
 -e^{-\Omega(x)} \frac{d}{dx} e^{\Omega(x)} =\frac{d}{dx}  +  W(x,\lambda)=A;
\qquad
 e^{\Omega(x)} \frac{d}{dx} e^{-\Omega(x)} =-\frac{d}{dx}  +
W(x,\lambda)=A^\dagger \label{EQ03}
 \end{equation}
and therefore
\begin{equation}
   H = A^\dagger A. \label{EQ042}
\end{equation}
The  SUSY partner of $H$ is obtained by simply changing the sign of the 
prepotential $\Omega$. The Hamiltonian written in this form is suitable for 
generalization to higher dimensions.

Working in three dimensions, a Hamiltonian for a
potential problem can be written as
\begin{equation}
     H_- = - e^{\Omega({\bf r})}\nabla e^{-2\Omega({\bf r})} \nabla 
              e^{\Omega({\bf r})},\label{EQ05}
\end{equation}
where $\nabla$ is the gradient operator in three dimensions. 
The partner Hamiltonian is defined by 
\begin{equation}
     H_+ = - e^{-\Omega({\bf r})}\nabla e^{2\Omega({\bf r})} \nabla 
              e^{-\Omega({\bf r})}. \label{EQ06}
\end{equation}
Again, we analyze the simplest case of the prepotential depending on a single
parameter $\lambda$ by means of the ansatz
\begin{equation}
         \Omega({\bf r},\lambda) = \lambda F({\bf r},\lambda).
\end{equation}
Proceeding as in the case of one dimension, the SUSY SI condition
leads to the Riccati equation for $F({\bf r}, \lambda)$
\begin{equation}
     |\nabla F({\bf r}, \lambda)|^2 -  \nabla^2 F({\bf r},\lambda) = K, 
\label{EQ08}
\end{equation}
where $K$ is a constant. This equation can be transformed into a free 
particle equation by writing  $F({\bf r},\lambda)= \log \chi({\bf r})$:
\begin{equation}
    \nabla^2 \chi({\bf r}) + K \chi({\bf r}) =0. \label{EQ092}
\end{equation}
Here again we will have three cases of $K >0, K=0 $ and  $K< 0$, 
separately. It is to be noted that, for a fixed 
constant $K$, in higher dimensions there are an infinite number 
of independent solutions, as against two solutions in one dimension. Hence the 
class of SIPs is already very large in higher dimensions
as compared to that in one dimension.

Now it is very easy to give examples of non-separable SIPs
in three dimensions, keeping in mind that any solution of free
particle Schr\"{o}dinger equation leads to a SIP. As an
example, consider a particular solution of \eqref{EQ092} for $K=0$, in the polar
coordinates
\begin{eqnarray}
\chi(r,\theta) &=&\sum_n\{ r^n  P_{n}(\cos(\theta)) +r^{-(n+1)}
P_{n}(\cos(\theta)) \},
\end{eqnarray}
with $\Omega(r,\theta,\lambda)=\lambda \log\chi(r,\theta)$. The corresponding
Hamiltonians $H_\pm$ become
\begin{equation}
   H_\pm = - \nabla ^2 + V^{(\pm)}(\vec{r}),
\end{equation}
with the  potentials $V^{(\pm)}(r)$ given by
\begin{equation}
   V^{(\pm)}(\vec{r}) = \lambda^2 (\nabla \chi)^2 \pm \lambda \nabla^2 \chi.
\end{equation}
Although we have found solutions to the SI requirement
for higher dimensional potentials, SI alone is insufficient  to
obtain  solutions for the energy eigenvalue problems. We also require
intertwining operators connecting the partner Hamiltonians.
A  straightforward extension of conventional intertwining relation  rapidly
becomes more and more complex as one takes up problems in higher dimensions and is likely to
become intractable. This suggests a need to look for a replacement for the
intertwining property of the partner Hamiltonians. The work on exact solutions of
potential problems within the path integral, mentioned in the introduction, suggests use
of reparametrisation of time, to be called space time transformation in this paper. 
A formulation of classical mechanics with an alternate parametrization of trajectories is well known and
has been used extensively for obtaining solutions of potential problems in one
and more than one dimensions within the path integral approach \cite{Duru} - \cite{HKlein}, \cite{cm1}, \cite{cm2}. 

For a classical problem defined by a Hamiltonian $H$,  the use of alternate
parametrisation of classical trajectories of energy $E$ leads to formulating the
classical problem in terms of a pseudo Hamiltonian $\Hsc= f(\mathbf{q})(H-E)$
where $f(\mathbf{q})$ is  a function of the generalized coordinates $\mathbf{q}$ 
and specifies the transformation to a 'new time parameter'. This motivates
an extension of SI requirement to more general Hamiltonians of the form
\begin{equation}
\Hsc= f({\mathbf q}) \Big(\frac{{\mathbf p}^2}{2m}  + (V({\mathbf
q})-E)\Big).\label{repar}
\end{equation}
The extension of SI to quantum Hamiltonians corresponding to  \eqref{repar} can
be done in several possible ways. Here we present  a particular scheme
and an example of application  to the radial equation coming from the
separation of variables in a spherically symmetric problem in three dimensions.

\paragraph*{Factorization and generalisations of shape invariance:}
The solution to the factorization problem \cite{infeld} of writing the
Schr\"{o}dinger
Hamiltonian
\begin{equation}
H= p^2 + V(x)-E
\end{equation}
in the factorized form $A^\dagger A$, where
\begin{equation}
    A = \frac{d}{dx} + Q(x), \qquad  A^\dagger = - \frac{d}{dx} + Q(x)
    \label{EQ041}
\end{equation}
is seen to be provided by the QHJ equation
\begin{equation}
  Q(x)^2 - \dd[Q(x)]{x} = V(x) - E.\label{EQ063}
\end{equation}
The solution $Q(x)$ of the Riccati equation \eqref{EQ041}
is equivalent to solving the Schr\"odinger  equation
\begin{equation}
    -\frac{d^2\psi(x)}{dx^2} +(V(x)-E) \psi(x) =0, \label{EQ02}
\end{equation}
since 
\begin{equation}
Q(x)= - \frac{\psi^{\prime}(x)}{\psi(x)}.
\end{equation}
Thus a solution of \eqref{EQ063} can be used to factorize the Hamiltonian $H$ as
$A^\dagger A$.

We now make use of this observation to extend shape invariance to more 
general situations which arise when one is trying to solve  a  Schr\"odinger
equation  by means of a point transformation. As a concrete  example we look for a function $Q(x)$
such that the operator appearing in a generalised form of the Schr\"odinger equation given by
\begin{equation}
  -\frac{1}{f(x)}\frac{d}{dx} f(x)\frac{d \psi(x)}{dx} + (V(x)-E)\psi(x)
=0.\label{EQ051}
\end{equation}
can be factorized and the above equation can be cast in the form
\begin{equation}
   \frac{1}{f(x)}   \left[- \Big(\frac{d}{dx}- Q(x)\Big) f(x) \Big(\frac{d}{dx}+
Q(x)\Big)\right]\psi(x)=0.\label{EQ393}
\end{equation}
 The requirement that \eqref{EQ051} and \eqref{EQ393} coincide, implies that
$Q(x)$ be a solution of the QHJ equation
\begin{equation}
  Q^2(x) -\frac{1}{f(x)}\frac{d}{dx}\Big(f(x) Q(x)\Big) = V(x) -E.\label{EQ071}
\end{equation}
Using factorization we can now define a SUSY partner equation
\begin{equation}
        \frac{1}{f(x)}\left[\Big(\frac{d}{dx}+ Q(x)\Big) f(x)
\Big(\frac{d}{dx}-
Q(x)\Big)\right]\psi(x) = 0. \label{EQ081}
\end{equation}
This then leads us to a generalization of SI property if we  demand that
the potentials $V_\pm$ obtained from 
\begin{equation}
 \frac{1}{f(x)}  A^\dagger f(x) A =  p^2 + V_-(x) -E,\label{EQ091}
\end{equation}   
be equal to  the potential $V^{(+)}(x)$ obtained from
\begin{equation}
  \frac{1}{f(x)} A f(x) A^\dagger =  p^2 + V_+(x) -E \label{EQ101}
\end{equation}
up to a constant and a redefinition of parameters in the usual fashion.
Using this extension of shape invariance property, we can apply methods of SUSY 
QM to a larger class of  equations. The partner potentials in the above 
equations can be written as
\begin{equation}
     V^{(\pm)}(x) = Q^2(x) \pm \frac{1}{f(x)}\frac{d}{dx}(f(x)Q(x)).
\end{equation}
Note that, for $f(x)=1$, the above expressions reduce to the forms familiar
from
the standard SUSY quantum mechanics.

In an alternative approach to define partner potentials and shape invariance, we
write the Hamiltonian given in \eqref{EQ051} as a product $CB$ where
\begin{equation}
    B= \frac{d}{dx} + Q(x), \qquad C = - \frac{d}{dx} -
\frac{f^{\prime}(x)}{f(x)} + Q(x)
\end{equation}
 and define the partner Hamiltonian as $H_+= BC$.  Expanding the products $BC$
and $CB$, the two Hamiltonians $H_\pm$ are seen to be
 \begin{eqnarray}
   H_\pm = -\frac{d^2}{dx^2} -\frac{f^{\prime}(x)}{f(x)} \frac{d}{dx} + V_\pm(x)
 \end{eqnarray}
 where the 'partner potentials' $V_\pm(x)$ are given by
 \begin{eqnarray}
   V^{(-)}(r) &=& Q^2(x) - Q^\prime(x) -Q(x)\Big(\frac{d\log f(x)}{dx}\Big),\\
   V^{(+)}(x) &=& Q^2(x) + Q^\prime(x) -Q(x) \Big(\frac{d\log f(x)}{dx}\Big)-
          \Big(\frac{d^2\log f(x)}{dx^2}\Big).
 \end{eqnarray}
\subsubsection*{An application to radial equation}
In general there are several possible ways of defining the partner potentials
and hence different ways of bringing in SI. Without loss of generality, we illustrate this by means of an example

Considering the radial equation in three dimensions,
\begin{eqnarray}
  -\frac{1}{r^2}\dd{r} r^2\dd[R(r)]{r} + V(r)  R(r)= E R(r)
\end{eqnarray}
which is of the form \eqref{EQ051}, if we make  replacements $x\to r$
and take $f(r)=r^2$, $Q=(\ell+1)/r$. The method described above can now be
used to find SIPs. These potentials will get related to
the solutions  of the radial equation instead of the free particle equation in one
dimension. Thus we will get a whole new class of shape invariant spherically
symmetric potentials.

In this case the partner potentials for the radial equation are
\begin{eqnarray}
  V^{(-)}(r) &=& Q^2(r) -\frac{1}{f(r)}\frac{d}{dr}(f(r)Q(r)),\\
         &=& \frac{\ell(\ell+1)}{r^2}.\\
  V^{(+)}(r) &=& Q^2(r) + \frac{dQ(r)}{dr} -Q(r) \frac{d\log f(r)}{dr}  -
          \frac{d^2\log f(r)}{dr^2}, \\
         &=& \frac{\ell(\ell-1)}{r^2}.
\end{eqnarray}
and the two Hamiltonians $H_\pm$ are
\begin{eqnarray}
   H_- &=& CB = - \DD{r} - \frac{2}{r} \dd{r}+\frac{\ell(\ell+1)}{r^2} \\
   H_+ &=& BC = - \DD{r} - \frac{2}{r} \dd{r} + \frac{\ell(\ell-1)}{r^2}
\end{eqnarray}
The operators $B$ and $C$,
\begin{equation}
    B = \dd{r} +\frac{\ell+1}{r}, \qquad C = \dd{r} + {2}{r}
\end{equation}
intertwine the solutions of the two Hamiltonians $H_\pm$.
Thus if $\psi^{(-)}_{\ell}(r)$ is an eigenfunction of $H_-$,
then
\begin{eqnarray}
      \psi^{(+)}_\ell(r) 
      &=&  B \psi^{(-)}_\ell(r)\\
      &=&r^{-(\ell+1)}\frac{d}{dr}\Big(r^{(\ell+1)}\psi^{(-)}_\ell(r)\Big),
\end{eqnarray}
is an eigenfunction of $H_+$. This is  just the relation between
the spherical Bessel functions $j_\ell(r), n_\ell(r), h_\ell(r)$ which are the
solutions for $H_-$ and $j_{\ell-1}(r), n_{\ell-1}(r), h_{\ell-1}(r)$, the
solutions  for $H_+$.\\

\section{Summary and Concluding Remarks}
In this paper we have analysed the shape invariance requirement and presented
an approach using a simple anstaz to reproduce all the know SI potentials in one
dimension. This approach to SI relates the superpotentials to solutions of free particle 
Schr\"{o}dinger equation and works equally well in higher dimensions.
Going beyond the simple ansatz of \eqref{WequaltoLF}, we arrive at the
isospectral shift deformation and to the  SIPs related to the EOPs \cite{isoshift}.
Further generalisations of the SI property are possible and are as follows.\\

\noindent
{\bf Generalizing isospectral shift}
The first generalization consists in replacing the constant, such
as that coming in shape invariance and isospectral shift conditions, by functions of $x$. 
Suppose we are give a potential $V(x)$. Solve QHJ and construct superpotential $W(x)$ for some energy $E$.
Then define 
\begin{equation}
    V^{(\pm)} = W(x)^2 - W^\prime(x).
\end{equation}
While carrying out isospectral shift we set $\tilde{W} = W(x)+ \phi(x)$ 
and try to find $\phi(x)$ by demanding that 
\begin{equation}
  \tilde{V}^{(+)}(x) = V^{(+)}(x) + R.
\end{equation}
The equation for $\phi(x)$ can be transformed into a second order 
Schr\"{o}dinger like equation. If that equation can be solved, this 
procedure will explicitly give a potential which is exactly solvable
and has its spectrum shifted by the constant $R$. This procedure can be 
repeated by replacing the constant $R$ by any function of $x$.
So for example we can take
\begin{equation}
  \tilde{V}^{(+)}(x) = V^{(+)}(x) + V_0(x)
\end{equation}
and set up an equation for $\phi(x)$. We can look for suitable $V_0(x)$
for which the equation for $\phi(x)$ can be solved. This generalization of SI
property, though known in literature, does not seem to have been investigated.
Using tools from QHJ and SUSYQM,  an investigation of applications of
this extension is under study and will be reported elsewhere.

\paragraph*{Generalizing shape invariance}
The second generalisation is indicated by re-examining  shape invariance
condition
\begin{equation}
     V^{(+)}(x, \lambda) = V^{(-)}(x,\rho(\lambda)) + K.
\end{equation}
We solved this equation by taking ansatz $w(x) = \lambda f(x)$.
We ended up with free particle equation. For each solution of 
the free particle equation, we then find a $W(x)$ and a shape invariant
potential. 

Again, if we replace the constant $K$ in the above condition by a function
$V_0(x)$, the same ansatz will lead to the Schr\"{o}dinger equation 
for potential $V_0(x)$. For each solution of Schr\"{o}dinger equation for 
this potential $V_0(x)$, we will be able to construct a potential which 
satisfies a generalized shape invariance condition
\begin{equation}
     V^{(+)}(x, \lambda) = V^{(-)}(x,\rho(\lambda)) + V_0(x).
\end{equation}

{\bf The second generalization of both the above processes to 
potentials in higher dimensions, can be obviously carried out 
without any difficulty.}\\

{\bf Acknowledgements:} The authors thank P K Panigrahi for useful discussions.
SSR acknowledges financial support from The Science \& Engineering Research
Board (SERB) / DST under the fast track scheme for young scientists (D.O. No:
SR/FTP/PS-13/2009). AKK thanks the faculty colleagues in the School of Natural Sciences  
for warm hospitality  extended to him during the one semester visit to the Shiv Nadar University, 
Noida, where a major part of this work was done and acknowledges the facilities provided
by the University. The authors are  thankful to researchers all over world who make their  manuscripts
available on the archives, http://xxx.arxiv.org/ and the people who maintain the archives.

\vspace{1cm}

\noindent
{\bf References}

\end{document}